\documentclass[12pt,aps,prl,draft,
showkeys,groupedaddress]{revtex4}
\usepackage{amsfonts}
\usepackage{theorem}
\usepackage{amsmath}
\usepackage{amssymb}

\def\Red{}
\def\Black{}
\def\Blue{}

\newcommand{\myfrac}[2]{
	\begin{array}{l}
	{\!{#1}\!}\\\hline
 	{\!{#2}\!}
	\end{array}\,
}
\newcommand{\der}[2]{
	\myfrac{\delta #1}{\delta #2}
}

\def\G{\Gamma}

\def\s{{\rm s}}
\def\S{{\cal S}}
\def\FP{Fadeev-Popoov}
\def\INT{\int\,\,d^4x\,}

\newcommand\QAPX[1]{{\Delta_{#1}(x)\cdot\G}\Black}
\newcommand\QCAP[1]{{\Delta_{#1}}\Black}
\newcommand\hepth[1]{{\tt hep-th/#1}}
\newtheorem{prop}{Proposition}
\newtheorem{lemma}{Lemma}
\newtheorem{definition}{Definition}
\newcommand{\props}[2]{    
        \begin{prop}{\label{prop:#1} #2}\end{prop}}

\begin{document}
\Red
\title{Non-Commutative Abelian Higgs model}
\Black

\author{Marco Picariello}
\affiliation{Universit\`a degli studi di Milano and INFN-Milano}
\Blue
\begin{abstract}
We investigate the Non-Commutative Abelian Higgs model.
We argue that it is possible to introduce a consistent renormalization
 method by imposing the Non-commutative BRST invariance of the theory
and by introducing the Non-Commutative Quasi-Classical Action Principle.
\end{abstract}
\Black
%
\maketitle
\section{Introduction}
Recently some efforts have been done to investigate the
 renormalizability of the Non-Commutative (NC) field
 theories~\cite{Picariello:2001mu,Barnich:2003wq,Amorim:2003wa,Brandt:2003fx,Martin:2002nr,RuizRuiz:2002hh,Aref'eva:1999sn,Chepelev:2000hm,Griguolo:2001wg,Petriello:2001mp}.
In this paper we consider a possible field theory candidate for the
 non-commutative extension of the Abelian Higgs model.
In particular we will show that the Non-Commutative Abelian Higgs
 ({\em NCAH}) model is stable under quantum
 corrections and then it is renormalizable in the sense
 of~\cite{Gomis:1995jp}.
In ordinary commutative theories, the main ingredient used to prove the
 renormalizability is given by the Quantum Action
 Principle~\cite{Lam:1973qa,Bergere:1975tr,Breitenlohner:1977hr,Clark:1976ym}.
As shown in~\cite{Stora2000,Ferrari:2002kz} the Quantum Action
 Principle (QAP), that affirms that
 the breaking of the Slavnov-Taylor is given by the insertion of a local operator,
 is a very strong requirement that can be relaxed into the Quasi-Classical Action Principle\ 
 (QCAP) in the proof of the renormalizability of
 theories.
The QCAP for general theories assumes that
 {\em the first non-vanishing order}
 in the loop expansion of the breaking of the Slavnov-Taylor
 is a local formal power series in the fields and external sources
 and their derivatives.
Under the assumption that the QCAP~\cite{Ferrari:2002kz} is valid
 for theories related to power-counting renormalizable theories 
 we can inferred, by using the Seiber-Witten (SW) map, that the the QCAP is a
 valid proprieties of non-commutative theories too.
We show that it is possible to formulate such theories, by using the
 Slavnov-Taylor identities (STi) related to the non-commutative extension of the gauge
 invariance.
In particular we will analyse the {\em NCAH model}, and we will study the
renormalization properties of it.
By using the NC extension of the QCAP and the SW map we find that
the {\em NCAH model}\ is renormalizable in the sense that it needs a finite set
of renormalization constant to become a well defined
theory~\cite{Gomis:1995jp}.
The {\em NCAH model}\ is as well predictable as the ordinary Abelian Higgs model.
The plan of the letter is the following:
First
 we show how to generalize the STi to NC theories;
then
 we introduce the meaning of the QCAP and
and we extent is to NC theories.
After that we show how, by using the STi and the
Non-Commutative Quasi-Classical Action Principle, it is
possible to study the renormalizability of the {\em NCAH model}.
%
%
\section{Slavnov-Taylor identities for Non-Commutative theories}\label{sec:STi}
Let us now give a look the the STi and let us generalize them to
NC theories.
First, as usually, given a fields transformation $s$:
\begin{eqnarray}
s \phi_i &=& R_{\alpha i}c_{\alpha}\,,
\end{eqnarray}
%
 where $R$ are generally formal power series of the fields and of their
 derivatives, and $c_{\alpha}$ are Grassmann fields which transforms,
 for example in the case of gauge symmetries, according to
\begin{eqnarray}
s c_a &=& -\frac{1}{2} f_{a}^{\ bc} c_b c_c\,,
\end{eqnarray}
%
 we couple the non linear transformation to additional external sources
 $\rho$ and add the new terms to the classical action.
Then the invariance can be written in a functional form as:
\begin{eqnarray}\label{eq:ST}
{\cal S}(\G^{(0)}) &=& \INT\left(
\der{\G^{(0)}}{\rho_i}\der{\G^{(0)}}{\phi_i}
 \right)= 0\,.
\end{eqnarray}
%
%
%
The real meaning of the functional derivative
 is that anywhere appear the field $Y$ one has to
substitute it with $X$.
This is an important subtlety when treating with non commutative
theories.
%
%
According to this full meaning
 we can
safely change the ordinary product into the $\star$-product and forget
about the previous meaning but use the operative standard rule: 
$X \star \der{}{Y}$ is for {\em left} functional derivatives, while 
$\der{}{Y}\star X$ is for {\em right} functional derivatives.
For these reasons, the NC extension of the STi in
eq.~(\ref{eq:ST}) is:
\begin{eqnarray}\label{eq:NCST}
{\cal S}(\G^{(0)\star}) &=& \INT\left(
\der{\G^{(0)\star}}{\rho_i}\star\der{\G^{(0)\star}}{\Phi_i}
 \right)= 0\,.
\end{eqnarray}
Operativally, we also remember the fundamental rules of an integrated
 $\star$-product~\cite{Szabo:2001kg,Douglas:2001ba}:
\begin{eqnarray}\label{eq:NCrot}
\INT a\star b \star c = (-1)^F\INT c \star a\star b\,,
\end{eqnarray}
 where $F$ is the fermi-bose factor, which take into account the fact
 that the fields are (anti)-commuting.
Once we have clarified the Slavnov-Taylor identities for a
 Non-Commutative theory, we need a further
step: the nilpotency of the linearized Slavnov-Taylor operator.
This is a  extension of the nilpotency of the ordinary linearized 
Slavnov-Taylor operator and it can be checked
 mechanically~\cite{Picariello:2003ky}, by using the definition in
 eq.~(\ref{eq:NCST}) and the properties of the $\star$-product.
\section{The Quasi-Classical Action Principle}\label{sec:QCAP}
It is customary to summarize the content of the QAP by characterizing
 the behavior of the renormalized quantum effective action $\G$  under
 infinitesimal variations of the fields and the parameters of the
 model.
For self consistency we report the full standard formulation
(as given for instance in~\cite{Ferrari:2002kz,Piguet:1995er}) in
 appendix~\ref{app:QAP}.
%
The QAP tells us that in a power-counting renormalizable theory
 the ST-like identity in eq.~(\ref{eq:QAP2})
 can be broken at quantum level
 only by the insertion of an integrated
 local composite operator of bounded dimension.
This is an all-order statement holding true regardless the normalization
 conditions chosen.
At the lowest non-vanishing order the insertion 
 reduces to a
 local polynomial in the fields and external sources and their derivatives
 with bounded dimensions.
This property is a consequence of the topological nature of the
 $\hbar$-expansion as a loop expansion.
That is, if a local insertion in the vertex functional
 were zero up to the order $n-1$, at the $n$-th order it must reduce
 from a diagrammatic point of view to a  set of points.
By power-counting this set is  finite and hence
 it corresponds to a local
 polynomial in the fields and the external sources and their
 derivatives.
The extension of the QAP beyond the power-counting renormalizable case
 is yet an open issue in the theory of renormalization.
In~\cite{Ferrari:2002kz} the QCAP have been introduced as a
consequence of the Stora conjecture~\cite{Stora2000}.
In the power-counting renormalizable case
 bounds on the dimensions can be given truncating the formal power series
 predicted by the QCAP to a local polynomial. Thus
 the QCAP reduces in this case to the
 part of the QAP stating that the lowest non-vanishing order
 $\Delta^{(n)}(x)$ of
 the breaking term 
 is a polynomial.
This justifies the name of QCAP.
For most practical purposes the QCAP (or, for power-counting renormalizable
 theories, the part of the QAP relevant to $\Delta^{(n)}(x)$) is what is
 really needed in order to carry out the program of Algebraic
 Renormalization.
In particular, this is enough to discuss the restoration of
anomaly-free ST-like identities order by order in the loop expansion.
This point have been illustrated in~\cite{Ferrari:2002kz} on the example
of the quantization of the Equivalence Theorem ST identities.
\subsection{The Non-Commutative theories}\label{sec:NCQCAP}
The existence of the SW map~\cite{Picariello:2001mu,Seiberg:1999vs},
which for example affirms that it is possible to construct a map that
relates an ordinary theory to a Non-Commutative one by conserving the gauge
symmetry properties, allows us to write an extension of the QCAP to NC
theories:
\props{NCQCAP}{
Let $\G^\star$ be the vertex functional corresponding to a Non-Commutative theory 
 with a classical action given by
$\G^{(0)\,\star}=\INT{\cal L^\star}(\Phi_a,\beta_i,\lambda)$
where $\Phi_a$ are the quantum fields,
 $\beta_i$ the external sources coupled to field polynomials
 $Q^i$, and $\lambda$ stands for the parameters of the model
 (masses, coupling constants, renormalization points).
Here the $^\star$ indicates that every fields multiplications are assumed
to enter in the Lagrangian via the NC product.
Notice that $\G^{(0)\,\star}$ is not power-counting
renormalizable however the commutative corresponding one, $\G^{(0)}$,
gives up to a power-counting renormalizable theory.
Let us define
\begin{eqnarray}
\S({\G^{\star}}) &=&\INT \left(
\alpha_a \der{\G^{\star}}{\Phi_a(x)}
+ \alpha_{ab}\Phi_b(x)\star \der{\G^{\star}}{\Phi_a(x)}
+ \alpha_{ia}\der{\G^{\star}}{\beta_i(x)}\star \der{\G^{\star}}{\Phi_a(x)}
+ \alpha\der{\G^{\star}}{\lambda}\right)\, .
\end{eqnarray}
Then {\rm (\bf Non-Commutative Quasi Classical Action Principle\rm)}
the first non-vanishing order in the loop expansion, say $n$,
of $\S(\G^{\star})$:
\begin{eqnarray}
\INT \Delta^{(n)\ \star} \equiv \S(\G^{\star})^{(n)} \, , ~~~~~
\S(\G^{\star})^{(j)}=0 \ \text{for } j=0,1,\dots,n-1
\label{eq:qcapbrkg}
\end{eqnarray}
is an integral of a formal power series in the fields and external
sources and their derivatives with given commutative bounded
dimensions but multiplied via the star product.
}
Notice that in our specific case, being the Slavnov-Taylor operator nilpotent,
the breaking term must satisfy the Wess-Zumino consistent condition: 
$\S(\INT \Delta^{(n)\ \star}) = 0$.
\section{The Non-Commutative Abelian Higgs model}
As it is well know, the Abelian Higgs model comes out form a global $O(2)$
 symmetric $\phi^4$ theory coupled to a local $U(1)$ Abelian gauge
 theory within the minimal coupling.
It is assumed that the potential for the complex $\phi$ field has the
 form of a Mexican hat and that the true perturbative minimum of the
 potential is obtained when the $\phi$ acquire a non trivial vacuum
 expectation value.
Written in term of the real and imaginary part of the $\phi$ the
 Lagrangian does not show any more the hidden global symmetry, however
 the UV properties are not modified by this dynamical symmetry
 breaking.
In particular the theory turns out to be power-counting renormalizable.
%
%
The Non-Commutative equivalent of the Abelian Higgs model
 can be obtained by substituting
everywhere in the Abelian Higgs model, in the unbroken phase, the ordinary product
with the $\star$-product.
Once the field $\phi$ acquires a vacuum expectation value, the action
become more involved, however it can be easily worked
out~\cite{Petriello:2001mp}.
By solving the Wess-Zumino consistent condition, it is possible to
show that the Non-Commutative Abelian Higgs model is renormalizable and that we need to introduce
the renormalization of the fields $A$, $\Phi$, $c$, and of the
constants $\mu$, $\lambda$, $g$ and of the tadpole $\nu$.
This is in accordance with the one loop computation done
in~\cite{Petriello:2001mp}.
\appendix
\section{The Quantum Action Principle}\label{app:QAP}
The standard formulation of the QAP is the following:

\props{QAP}{
Let $\G$ be the vertex functional corresponding to a
 (power-counting renormalizable) theory in a $D$-dimensional
 space-time with a classical action given by
\begin{eqnarray}
\G^{(0)}=\int\,d^Dx\, {\cal L}(\varphi_a,\beta_i,\lambda)
\label{eq:classact}
\end{eqnarray}
where $\varphi_a$ are  the quantum fields,
 $\beta_i$ the external sources coupled to field polynomials
 $Q^i$ and $\lambda$ stands for the parameters of the model
 (masses, coupling constants, renormalization points).
Let the inverse of the quadratic part of the action be the standard
 Feynman propagators.\\
Given the local operator
\begin{eqnarray}\label{eq:QAP0}
\S({\G}) \equiv
\alpha_a \der{\G}{\varphi_a(x)}
+ \alpha_{ab}\varphi_b(x) \der{\G}{\varphi_a(x)}
+ \alpha_{ia}\der{\G}{\beta_i(x)}\der{\G}{\varphi_a(x)}
+ \alpha\der{\G}{\lambda},
\end{eqnarray}
 where $\alpha_a, \alpha_{ab}, \alpha_{ia}$ and $\alpha$ are constants,
 then the Quantum Action Principle can be stated in the
 following way
\begin{eqnarray}\label{eq:QAP1}
\S({\G})= \QAPX{ } = \Delta^{(n)}(x)
+ O(\hbar^{n+1}).
\end{eqnarray}
 $\QAPX{ }$ denotes the insertion of a local operator.
 Moreover the lowest non-vanishing order coefficient $\Delta^{(n)}(x)$
 of $\QAPX{ }$ is a local
 polynomial in the fields  and external sources and their derivatives
 with bounded dimension.
}
At the integrated level (Slavnov-Taylor-like identities) the QAP
 reads
\begin{eqnarray}
{\cal S}(\G) \equiv \INT\S(\G) = \INT\QAPX{ } \, .
\label{eq:QAP2}
\end{eqnarray}
The first non-vanishing order of the ST-like breaking terms is
 given by
\begin{eqnarray}
\Delta^{(n)} \equiv {\cal S}(\G)^{(n)} = \INT\Delta^{(n)}(x) \, .
\label{eq:intSTI}
\end{eqnarray}
As a consequence of Proposition~\ref{prop:QAP}, $\Delta^{(n)}$
 is an integrated local polynomial in the fields and the external
 sources and their derivatives with bounded dimension.
The ultraviolet (UV) dimension $d_\QCAP{ }$ of $\QCAP{ }^{(n)}$ can be
 predicted from the UV dimensions
 $d_a$ of the fields $\varphi_a$ and the UV dimensions $d_{Q^i}$ of
 the field polynomials $Q^i$~\cite{Piguet:1995er}.
We do not dwell on this problem here
 since the only information we need for the present discussion is the
 fact that  $d_\QCAP{ }$ is bounded.

\end{document}